\documentstyle[11pt,twoside,graphicx,jltp]{article}

\title{Comparison of energy and phase relaxation in metallic wires}

\author{A.B. Gougam\address{Department of Physics and Astronomy, 
Michigan State University, East Lansing, MI 48824-1116, USA}, F. Pierre$^*$, 
H. Pothier$^*$, D. Esteve\address{Service de Physique de l'Etat Condens\'{e}, 
Commissariat $\grave{a}$ l'Energie Atomique, Saclay, F-91191 Gif-sur-Yvette, France}
\linebreak and Norman O. Birge}

\runninghead{A.B. Gougam {\it et al.}}{Comparison of energy and phase
relaxation in metallic wires}
       
\begin{document}

\begin{abstract}
We have determined the phase coherence time, $\tau_{\phi}$, from
magnetoresistance measurements of long, narrow
wires of Au, Ag, and Cu, over the temperature range 40 mK-6 K. In the Cu
and Au wires, $\tau_{\phi}$ saturates at low temperature. In the Ag wire,
$\tau_{\phi}$ continues to increase down to the lowest temperatures measured;
moreover, its temperature dependence below about 1 K is consistent with 
theoretical predictions of Altshuler, Aronov and Khmelnitskii published
in 1982. These results cast doubt on recent assertions that
saturation of $\tau_{\phi}$ at low temperature is a universal phenomenon
in metal wires. We compare these results with those of recent experiments
on energy relaxation in similar metallic wires. The results of the two 
experiments are strongly correlated, suggesting that a single (unknown)
mechanism is the cause of the enhanced phase and energy relaxation observed
in some samples.

PACS numbers: 73.23.-b, 73.50.-h, 71.10.Ay, 72.70.+m.

\end{abstract}

\maketitle

%Include this space if you do not use sections in your document.
%\vspace{0.3in}

\section{INTRODUCTION}
One of the fundamental properties of disordered conductors that has 
been studied extensively the past 20 years is the phase coherence 
time of the conduction electrons. At temperatures below about 1 K, 
the dominant phase-breaking mechanism in nonmagnetic disordered 
metals is predicted to be electron-electron scattering.\cite{AAK,review} 
The corresponding phase-breaking rate decreases as the temperature 
is lowered as a power law, $\tau_{\phi}^{-1}$ $\propto$ $T^p$, where
$p$ depends on the effective dimensionality of the system. 
Recent experiments and analysis of several older 
experiments\cite{PM} show that the phase coherence time in 
some samples tends to saturate at a finite temperature, in contrast
with theoretical expectations. A second consequence of electron-electron
scattering is energy exchange between quasiparticles. Recently, 
experiments have been performed which provide information on the energy 
dependence, as well as the overall rate, of energy exchange in mesoscopic
wires driven out of equilibrium by a current.\cite{HP1,HP2} Those 
experiments showed that the energy relaxation rate in Cu wires was 
larger than predicted by theory, and the energy dependence of the scattering
rate was different from that predicted.

The purpose of the present work is twofold. First, we aim to shed light
on the issue of saturation of 
the phase coherence time, by measuring the temperature dependence of 
$\tau_{\phi}$ in wires of several different metals. Second, we compare
our results with those from the energy relaxation experiments, to see
if energy and phase relaxation are related in similarly prepared samples.
Energy relaxation experiments have now been performed on wires of Ag 
\cite{FP} and Au\cite{UFP} in addition to the original work on Cu. For purposes 
of comparison, we have therefore studied phase coherence in Cu, Ag,
and Au wires.

\section{EXPERIMENT}
The wires used in our experiment were deposited on similar substrates,
and in the same
electron-gun evaporator used to fabricate samples for the energy 
relaxation experiments.\cite{HP2,FP} The wires were patterned using e-beam
lithography and the lift-off process. Sample dimensions are given in 
Table I. The three samples have very similar resistivity $\rho$, as deduced from
the resistance $R$, length $L$, width $w$ and thickness $t$. 
The diffusion constants $D$ were deduced from $\rho$ using Einstein's relation.
The only difference in the fabrication procedure for these
three samples is that the Au sample was deposited on top of a thin (1 nm)
layer of Al to improve adhesion to the substrate. Sample widths vary 
from 65 to 110 nm, hence all the samples are quasi one-dimensional with
respect to the phase-breaking length, $L_{\phi}=\sqrt {D\tau_{\phi}}$
and the thermal length,
$L_T=\sqrt {\hbar D/k_B T}$ over the temperature range studied.  
The samples were immersed in the dilute phase of the $^3$He-$^4$He mixture
of a dilution refrigerator. Electrical lines to the sample were filtered
at the top of the cryostat and again near the sample. Resistance 
measurements were performed using a standard ac four-terminal technique 
with a lock-in amplifier. A ratio transformer was used in a bridge 
circuit to enhance the measurement sensitivity to small changes in 
sample resistance.
 
Figure 1 shows the magnetoresistance of the Ag sample at several temperatures.
At low temperature, the magnetoresistance is positive, indicating that the spin-orbit
scattering length $L_{so} = \sqrt {D\tau_{so}}$ is shorter than $L_{\phi}$. 
Figure 1 shows that the magnitude of the 
weak localization contribution to the magnetoresistance continues to increase
down to the lowest temperatures measured. Figure 2 shows similar raw data for 
the Cu sample. In contrast to the Ag data, the magnitude of the magnetoresistance
does not continue to grow at low temperature, but saturates at about 1 K in 
the Cu sample. 
\begin{table}
\label{Table I}
\centerline {\small\bf{TABLE I}}
\vspace{0.2in}
%\caption{}
                                                                                                                                                                                                                                                                                                                                                                                                                                                                                                                                                                                                                                                                                                                                                                                                                                                                                                                                                                                                                                                                                                                                                                                                                                                                                                                                                                                                                                                                                                                                                                                                                                                                                                                                                                                                                                                                                                                                                                                                                                                                                                                                                                                                                                                                                                                                                                                                                                                                           
\begin{tabular}{|l|ccccccc|}
\hline
Sample    &  $L$  & $w$  & $t$    & $R$  & $\rho$ & $D$ & $L_{so}$ \\
material  &($\mu$m)&(nm)&(nm)&(k$\Omega$)&($\Omega\cdot$nm)&($\rm{m}^2 \cdot \rm{s}^{-1}$)&(nm) \\
\hline
Au  &  271     &   100   &  45  &  2.0  & 33  & 0.010 & 58.5\\ 
Ag  &  136     &   65    &  45  &  1.4  & 30  & 0.012 & 75 \\
Cu  &  271     &   110   &  45  &  1.9  & 35  & 0.007 & 520 \\ 
\hline
\end{tabular} 
\end{table}

\begin{figure}[!tb]
  \centering
  \includegraphics[angle=0,width=3.5in,height=4.0in]{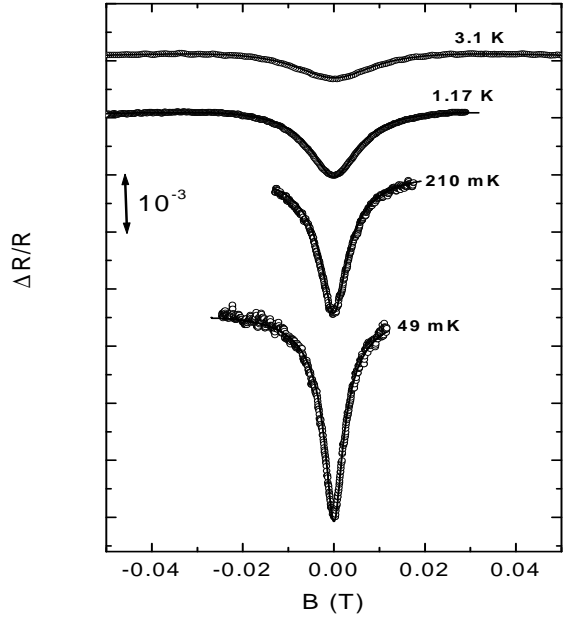}
  \caption{Magnetoresistance data (circles) and fits to Eq. 1 (solid lines)
     for Ag sample at different temperatures. The curves are offset 
     vertically for clarity.}
  \label{Figure 1}
\end{figure}

\begin{figure}[!tb]
  \centering
  \includegraphics[angle=0,width=3.1in,height=4.0in]{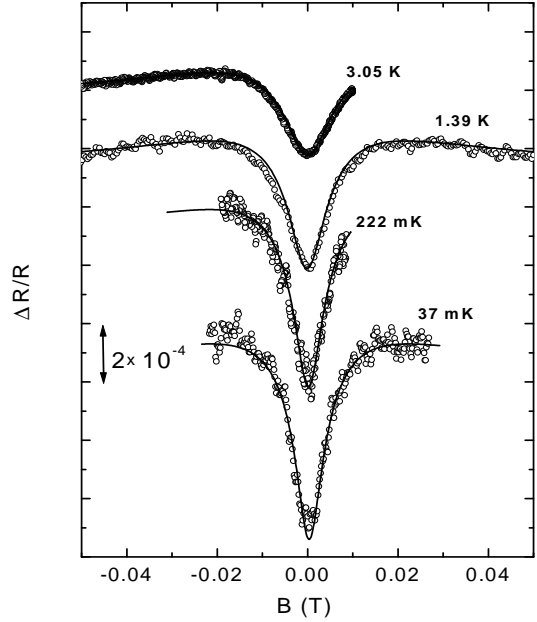}
  \caption{Magnetoresistance data (circles) and fits to Eq.\ 1 (solid lines)
     for Cu sample at different temperatures. The curves are offset 
     vertically for clarity. Note the different vertical scales in Figs. 1 and 2.}
  \label{Figure 2}
\end{figure}

The magnetoresistance is very small in the Au sample, indicating that 
$L_{\phi}$ is very short. In fact, the poor signal-to-noise ratio of
the Au data prevented us from obtaining reliable data at temperatures
below 1 K. The data we were able to obtain indicate that $L_{\phi}$
had already saturated at 6 K.

\section{ANALYSIS AND DISCUSSION}
To obtain estimates of $L_{\phi}$ and hence
$\tau_{\phi}$, we fit our data to the weak localization expression for the 
magnetoresistance in 1D:\cite{AA}
\begin{eqnarray}
{{\Delta}R \over {R}}& = &{{e^{2}R} \over {\pi \hbar L}} \left\{ {{3} \over {2}} 
\left[{{1} \over {L_\phi ^2}} + {{4} \over {3L_{so}^2}} + 
{{1} \over {3}}\left({{wBe} \over {\hbar}}\right)^2\right]^{-1/2} \right . \\ \nonumber
&  & \left . -{{1} \over {2}} \left[{{1} \over {L_\phi ^2}} + 
{{1} \over {3}}\left({{wBe} \over {\hbar}}\right)^2\right]^{-1/2} \right\}
\end{eqnarray}
where we have omitted terms that describe spin-flip 
scattering by magnetic impurities. Reliable determination of 
$L_{so}$ from the fit is only possible 
at the higher temperatures shown, where the magnetoresistance 
changes slope at high fields.
At lower temperatures, we fix the value of $L_{so}$ and fit to 
the single free parameter $L_{\phi}$. 

Echternach et al.\cite{Echternach} have emphasized that Eq.\ (1) is not
strictly valid at temperatures
where the dominant phase-breaking mechanism is electron-electron 
scattering with small energy
transfer, also called Nyquist scattering.\cite{AAK} In that 
low-temperature limit, phase 
differences accumulate slowly over many collisions, hence 
dephasing is not an exponential process.
We refer the reader to Ref. 9 for a complete discussion 
of this issue.
Here it suffices to say that experimental data does not allow one to distinguish
between the correct magnetoresistance formula in the Nyquist regime and Eq.
(1). The reason is that the function that enters in the Nyquist regime,
$f(x)=Ai(x)/Ai^{\prime }(x),$ with $Ai$ the Airy function, is well approximated
for all $x$ by $-(0.5+x)^{-1/2}$, with corrections of at most 4\%. 
Use of this approximation in the correct magnetoresistance 
formula\cite{corrections} results in
Eq. (1) with $\tau _{\phi}^{-1}=0.5\tau _{N}^{-1}+\tau _{\phi 0}^{-1}$,
where $\tau _{N}^{-1}$ is the Nyquist dephasing rate and $\tau _{\phi 0}^{-1}$
is the dephasing rate due to electron-phonon scattering
and any other mechanisms.
Realistically, $\tau _{\phi}$ defined in this way is the only time
accessible to experiment.

Figure 3 shows $\tau_{\phi}$ versus temperature for our three samples. The differences, 
already apparent in the raw data, are striking. In the Ag sample, $\tau_{\phi}$ 
continues to increase down to the lowest temperatures measured. In the Cu sample, 
$\tau_{\phi}$ saturates below 1 K at a value of 2 ns, while in the Au sample,
$\tau_{\phi}$ saturates already at 6 K at a value of 10 ps.
\begin{figure}[!tb]
  \centering
  \includegraphics[angle=0,width=3.2in,height=4.2in]{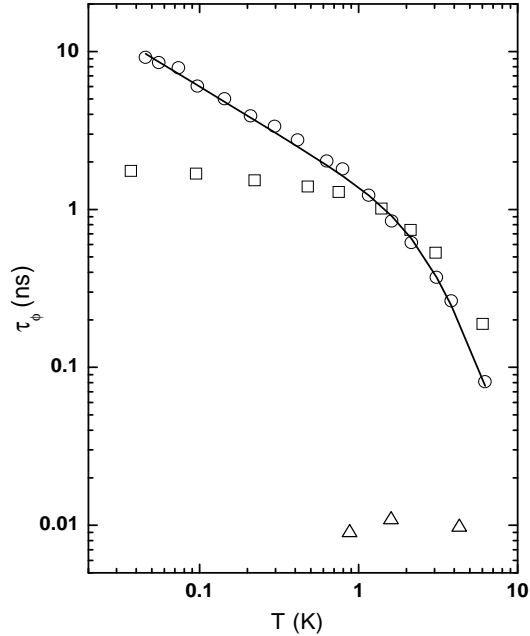}
  \caption{Phase coherence time for the Ag($\circ$), Cu($\Box$) 
    and Au($\triangle$) samples as a function of temperature.  The solid line
    is a fit of the Ag data to Eq.\ (2).}
  \label{Figure 3}
\end{figure}
The theory of electron-electron scattering in 
disordered conductors predicts that $\tau_{\phi}$ should increase as a power law
as the temperature is lowered. For the 1D case, the predicted 
behavior in the Nyquist regime\cite{AAK,corrections} 
is $\tau_{\phi}=2\tau _{N}=2\left( \frac{e^{2}R}{\hbar L}%
\frac{k_{B}T\sqrt{D}}{\hbar }\right) ^{-2/3}$.
To see if our Ag data is consistent with this prediction, we have fit the 
temperature dependence of $\tau_{\phi}$ to the functional form:
\begin{equation}
\tau_{\phi}^{-1}=AT^{p}+BT^{3}
\end{equation} 
where the second term is meant to describe electron-phonon scattering at higher 
temperatures.\cite{FT1} The best values of the three fit parameters are 
$p$=0.61, $A$=0.7 ns$^{-1}$K$^{-0.61}$, and $B$=55 $\mu$s$^{-1}$K$^{-3}$, 
compared to the theoretical predictions $p$=2/3,  
$A$=0.6 ns$^{-1}$K$^{-2/3}$. 
Thus the Ag data are in agreement with the 
theory of Nyquist dephasing in both the overall magnitude and temperature 
dependence.\cite{FT2} 
These results are consistent with those of Wind et al. \cite{Wind},
who measured both the magnitude and width dependence of the Nyquist 
dephasing rate in Ag and Al wires above 2 K, 
and with those of Echternach and al. \cite{Echternach}, who
found agreement with theory for the dephasing rate of Au samples down to 100
mK.  Our results thus extend by a factor of two lower temperature
the experimental confirmation of the Nyquist scattering theory.

What conclusions can we draw from these results? First, our Ag data contradict
the recent experimental claim that saturation of $\tau_{\phi}$ is universal in
disordered metal wires.\cite{PM} The proposed theoretical expression for the maximum
value of $\tau_{\phi}$ presented in that paper gives a result $\tau_{\phi max}=2$ ns 
for our Ag sample, about 5 times shorter than what we observe at our lowest 
temperature. Second, since the macroscopic
parameters (physical dimensions and resistivity) of our three
samples are nearly identical,
{\underline {any}} theoretical model that predicts a maximum value of 
$\tau_{\phi}$ based on those 
parameters alone must be incorrect.\cite{GZ} Our work thus lends
experimental support to recent theoretical papers\cite{AAG,IFS} refuting 
the claim that saturation of $\tau_{\phi}$ is universal. 
On the other hand, our Cu and Au data show that {\underline {some}} samples
do show saturation of $\tau_{\phi}$. Since all three of our samples were measured
in the same cryostat under the same conditions, it is unlikely that the 
saturation is due to interference from external electromagnetic radiation, as 
was recently proposed.\cite{AAG} The cause of the saturation is not yet known,
but we will mention some recent proposals at the conclusion of this paper. 
The small value of $\tau_{\phi}$ in our 
Au sample is curious, when compared with data from other
workers.\cite{PM,Echternach} Further measurements on other samples will reveal 
whether the Al underlayer has any influence on the dephasing rate.
\section{COMPARISON WITH ENERGY RELAXATION}
We now compare these results with those from the energy relaxation experiments.
\cite{HP1,HP2,FP,UFP} The analysis of those experiments was performed within the 
framework of the quantum Boltzmann equation.\cite{N,KR} The results of the 
analysis were expressed in terms of a kernel function, $K({\varepsilon})$, 
which describes the scattering rate between quasiparticles as a function of 
the energy $\varepsilon$ exchanged in the interaction. In the Appendix, 
we present a heuristic
argument relating $K({\varepsilon})$ to the dephasing rate ${\tau_{\phi}}^{-1}$. 
The main result of that calculation is that if 
$K({\varepsilon})\propto {{\varepsilon}^{-{\alpha}}}$
with ${\alpha}>1$, then the dephasing rate should have the temperature dependence 
$\tau_{\phi}^{-1} \propto T^{1/\alpha}$. 
Starting with Ag, Pierre et al.\cite{FP} have found that with ${\alpha} \approx 1.5$ in
some samples, and ${\alpha}=1.2$ in others. The former result is in agreement
with the theory of electron-electron interactions,\cite{review} and is consistent
with our observation that ${\tau_{\phi}}^{-1} \propto {T^{0.61}}$ at low temperatures.
Furthermore, the prefactor in $K({\varepsilon})$ is close to the theoretical prediction,
consistent with our similar observation for the dephasing rate. In Cu and Au,
the situation is more subtle. The original energy relaxation experiments\cite{HP2}
in Cu found $K(\varepsilon)={\tau_0^{-1}}{\varepsilon^{-2}}$, 
with $\tau_0 \approx 1$ ns.
 The characteristic scattering time of $1$ ns is very close to the
dephasing time we observe at low temperature, 
but the energy dependence of $K(\varepsilon)$ is not consistent
with our observation of $\tau_{\phi}$ saturation.
According to the derivation given in the Appendix, 
$K(\varepsilon) \propto \varepsilon^{-2}$ should lead to 
${\tau_{\phi}}^{-1}\propto {T^{1/2}}$.  In Au, the 
situation is similar. The energy relaxation experiments\cite{UFP} find  
$K(\varepsilon)={\tau_0^{-1}}{\varepsilon^{-2}}$,
with $\tau_0=100$ ps, whereas we observe saturation of $\tau_{\phi}$ 
at 10 ps.  We note that the Au samples used in the energy 
relaxation experiment did not have the thin Al underlayer.

The fact that the energy dependence of $K(\varepsilon)$ deduced from the energy relaxation
experiments in Cu and Au is not consistent with the observed temperature dependence
of $\tau_{\phi}$ may simply be telling us that electron-electron collisions are not responsible
for one or both of those observations. The analysis of the energy relaxation experiments
assumes that two-body electron-electron collisions are the dominant energy exchange 
process in the sample. The relation between $K(\varepsilon)$ and $\tau_{\phi}$ presented
in the Appendix assumes the same for dephasing. Since the results of neither 
experiment agree with the theoretical prediction for electron-electron scattering,
it is perhaps not surprising that our comparison of the results from within that
framework leads to a contradiction.

\section{CONCLUSIONS}
What then is the cause of the saturation of $\tau_{\phi}$ we observe in our Cu and Au 
samples? It was already argued by Mohanty et al. that magnetic impurities are 
unlikely to be the cause of the saturation, since samples with magnetic impurities 
deliberately added do not exhibit a saturation of the dephasing rate.\cite{PM} 
Our results support that conclusion. Although it is well known that magnetic impurities 
can lead to dephasing via spin-flip scattering, it seems unlikely that
they could provide an efficient mechanism of energy exchange 
in the absence of an external magnetic field. There have been recent suggestions 
that two-level systems may be responsible.\cite{IFS,ZDR} One of those 
proposals\cite{IFS}  
relates the dephasing rate to the level of $1/f$ noise in the sample at frequencies close 
to $\tau_{\phi}^{-1}$. That proposal relies on assumptions about the 
distribution of tunneling centers
in disordered metals that have not been directly tested by experiment. The second 
proposal invokes two-level systems with nearly degenerate ground states, which act
as two-channel Kondo impurities.\cite{ZDR} It is argued that such systems can lead
to a temperature-independent dephasing rate over a limited temperature range, 
below which the rate must tend to zero.  It remains to be seen if either of these
ideas can explain the intriguing results regarding both phase and energy
relaxation in metal wires. 

\section*{ACKNOWLEDGMENTS}
We are grateful to M. Devoret and Y. Imry for valuable discussions. This work was
supported in part by NSF grant DMR-9801841. Travel between Saclay and East Lansing
was supported by NATO, under Collaborative Research Grant 970273.

\section*{APPENDIX}
To facilitate comparison of the energy relaxation and phase relaxation experiments, 
we present a heuristic derivation of the dephasing rate one obtains from a given 
$K({\varepsilon})$.\cite{I} The out-scattering term in the Boltzmann 
equation takes the form:
\begin{equation} 
I^{out}(E,\{ f \})=\int{d{\varepsilon}dE'K({\varepsilon})
f(E)[1-f(E-{\varepsilon})]f(E')[1-f({\varepsilon}+E')]}
\end{equation}
where we have suppressed all reference to spatial variables. We assume that the 
dephasing rate at temperature $T$ is equal to the out-scattering rate with the 
initial state occupied and with equilibrium Fermi-Dirac distribution functions $f_{FD}$. 
Within the framework of Eq.\ (3), it makes no sense to include scattering events 
with energy transfer less than the dephasing rate itself, hence we set the lower 
limit of the integral to ${\hbar}/{\tau_{\phi}}$.\cite{I} The upper limit is cut off 
at $\approx k_BT$ by the availability of unoccupied final states. We then have:

\begin{equation}
\tau_{\phi}^{-1} \approx \int_{{\hbar/\tau_{\phi}}}^{k_BT}{d{\varepsilon}
K({\varepsilon})g({\varepsilon})}
\end{equation}
where 

\begin{equation}
g({\varepsilon})=\int {dE^{'}f_{FD}(E^{'})[1-f_{FD}(E^{'}+{\varepsilon})]}=
{\varepsilon}[1-e^{-{\varepsilon}/{k_BT}}]^{-1}
\end{equation}
For $0\leq {\varepsilon} \leq k_BT$, $g({\varepsilon}) \approx k_BT$, so 
the final result is:

\begin{equation}
\tau_{\phi}^{-1} \approx k_BT \int_{{\hbar/\tau_{\phi}}}^{k_BT}
{d{\varepsilon}K({\varepsilon})}
\end{equation}
If $ K({\varepsilon}) \propto {\varepsilon}^{-{\alpha}}$ with ${\alpha}>1$, 
then the integral 
is dominated by the low-energy limit, and the dephasing rate has the temperature 
dependence ${\tau_{\phi}}^{-1} \propto T^{1/\alpha}$. For the case of disordered 
metals of dimension d, the theoretical prediction\cite{review} is  
$K({\varepsilon}) \propto {\varepsilon}^{(d-4)/2}$. In 1D, this leads
to the well-known result\cite{AAK} ${\tau_{\phi}}^{-1} \propto T^{2/3}$.

We note here some possible limitations of the approach taken above.  Most notably, 
Eq.\ (4) indicates that dephasing occurs by single scattering events with energy 
transfer greater than $\hbar/\tau_{\phi}$. This is in contrast with the discussion 
given in the original theoretical papers,\cite{AAK} where dephasing is described 
by a process 
of gradual accumulation of phase during many collisions with energy transfer 
$< \hbar/\tau_{\phi}$. Second, the approach shown here does not address 
subtle differences between dephasing rates measured in different experiments. 
Recently it has been shown both theoretically \cite{B} and experimentally \cite{HCB} 
that the dephasing rate measured in weak localization experiments is different from 
that measured in universal conductance fluctuation experiments. Nevertheless, 
we find it remarkable that the simple argument outlined above gives the correct 
temperature dependence of the dephasing rate from the energy dependence of 
$K({\varepsilon})$.


\begin{thebibliography}{9}
\bibitem{FT1} The data presented here do not extend to high enough temperature to 
distinguish the $T^3$ law from other powers.  The $T^3$ term has been
used to describe dephasing above 1 K in similar samples in Refs. 3 and 9, 
among others.
\bibitem{FT2} During the course of the Ag experiment, we 
slightly underestimated the effect of heating due to the drive current. Since 
electron cooling at the lowest temperatures occurs almost entirely by thermal 
conduction through the leads, one can calculate the effect of heating using the 
Weidemann-Franz relation and the known sample geometry. The drive current at the 
lowest three temperatures shown was 5 nA, which at 50 mK causes a temperature rise 
of about 5 mK at the center of the wire.  Use of a lower drive current might 
result in a slightly larger value of p from the fit.  This error was corrected in 
the Cu experiment, where electron heating due to the drive current was kept below 
a few percent of the lattice temperature.
\end{thebibliography}
\end{document}